# Efficacy of Electrically-Polarized 3D Printed Graphene-blended Spacers on the Flux Enhancement and Scaling Resistance of Water Filtration Membranes

Numan Yanar[a], Hosik Park[b], Moon Son[c*], Heechul Choi[a*]

[a] School of Earth Sciences and Environmental Engineering, Gwangju Institute of Science and Technology (GIST), 123-Cheomdangwagi-ro, Buk-gu, Gwangju, Republic of Korea

[b] Green Carbon Research Center, Chemical Process Division, Korea Research Institute of Chemical Technology (KRICT), Daejeon 34114, Republic of Korea

[c] School of Urban and Environmental Engineering, Ulsan National Institute of Science and Technology, 50, UNIST-gil, Eonyang-eup, Ulju-gun, Ulsan, 44919, Republic of Korea

---

[*] Co-corresponding authors.

E-mail addresses: hcchoi@gist.ac.kr (Heechul Choi)

moonson619@unist.ac.kr (Moon Son)

Phone: +82-62-715-2441; Fax: +82-62-715-2423.

**Abstract**

In this research, an electrically-polarized graphene-polylactic acid (E-GRP) spacer is introduced for the first time by a novel fabrication method, which consists of 3D printing followed by electrical polarization under a high voltage electric field (1.5 kV/cm). The fabricated E-GRP was tested in an osmotic-driven process (forward osmosis system) to evaluate its performance in terms of water flux, reverse solute flux, and ion attraction compared to a 3D printed non-polarized graphene-polylactic acid (GRP) spacer and a polylactic acid (PLA) spacer. The use of the developed E-GRP spacer showed > 50% water flux enhancement (32.4 ± 2 LMH) compared to the system employing the GRP (20.5 ± 2.3 LMH) or PLA (20.8 ± 2.1 LMH) spacer. This increased water flux was attributed to the increased osmotic pressure across the membrane due to the ions adsorbed on the polarized (E-GRP) spacer. The E-GRP spacer also retarded the gypsum scaling on the membrane compared to the GRP spacer due to the dispersion effect of electrostatic forces between the gypsum aggregation and negatively charged surfaces. The electric polarization of the E-GRP spacer was shown to be maintained for > 100 h by observing its salt adsorption properties (in a 3 M NaCl solution).

**Introduction**

Water scarcity has been a critical problem in the last few decades due to drought accelerated by global warming (Mekonnen and Hoekstra 2016). As fresh water resources are already limited, the need for treating non-potable waters has emerged (Okun 1996). As of 2020, we are confronting a global pandemic and its indirect side effects, one of which is the mixing of viruses with water (Annalaura, Ileana et al. 2020, Bhowmick, Dhar et al. 2020, La Rosa, Bonadonna et al. 2020). Although the United States Environmental Protection Agency stated that drinking water is still safe for drinking (U.S. Environmental Protection Agency 2020), Casanova et al. stated that coronaviruses can remain infectious in water for days to weeks (Casanova, Rutala et al. 2009). Since we are faced with this critical situation that is endangering human lives, virus-selective water treatment methods come to the front compared to other treatment methods. When virus-selectivity is a concern, membrane-treatment methods can be considered the primary approach due to their excellent selectivity of water over other matter (Naddeo and Liu 2020). For the success of the membrane process to secure clean water, the membrane's performance should be maintained with minimal changes in energy consumption over long-term operation; however, most studies have focused on the development of novel membranes (Fan, Harris et al. 2001, Nguyen, Tobiason et al. 2011) and are already at their limit for the current technology (Rastogi 2020). In addition, other parts of the membrane processes should also be further developed to increase the overall efficiency of the membranes processes.

The development of the channel spacer is one of the most critical parts affecting the performance of the membrane process as it distributes the feed water stream and works as a protective guard to the membrane surface (Abid, Johnson et al. 2017). To date, most studies on spacers have been conducted by computer modeling with a focus on the shape of the spacers (Rastogi 2020). There are relatively few investigations using in-situ filtration tests, although computational fluid dynamics (CFD) modelling of the geometry has been extensively performed (Siddiqui, Farhat et al. 2016, Kerdi, Qamar et al. 2018, Ali, Qamar et al. 2019, Yanar, Kallem et al. 2020).

The fabrication material, surface structure, and functionality of spacers also carry great importance. In this regard, Yanar et al. compared the performances of acrylonitrile butadiene styrene (ABS), polylactic acid (PLA), and polypropylene (PP) as the spacer materials by employing them in a forward osmosis (FO) type membrane filtration system and clearly observed differences in fouling and ion attraction by the spacers (Yanar, Son et al. 2018). On the other hand, there have been a lot of studies on different types of coatings for spacers to enhance their surface properties. Some examples of such studies are: a nanosilver surface modification to a reverse osmosis (RO) membrane and a spacer for mitigating biofouling (Yang, Lin et al. 2009), an antibacterial spacer obtained by the sonochemical deposition of ZnO nanoparticles (Ronen, Semiat et al. 2013), nanosilver-modified feed spacers for anti-biofouling of ultrafiltration (UF) membranes (Ronen, Lerman et al. 2015), a polyaniline and polypyrrole (PPy) coating onto a stainless steel grid for high selectivity oil-water separation (An, Cui et al. 2014), graphene coating on a steel mesh for oil/organic-water separation (Sun, Li et al. 2013), and the formation of nanotubes on stainless steel meshes by electropolymerization (Darmanin and Guittard 2016). Nevertheless, surface coating has not always been successful at enhancing the spacer performance. For example, polydopamine, polydopamine-g-PEG, and copper coatings on feed spacers did not show a high performance in mitigating biofouling (Araújo, Miller et al. 2012, Miller, Araújo et al. 2012). Similarly, CuO modified spacers also did not improve the performance of RO systems in terms of biofouling (Yang, Son et al. 2019).

Alternative techniques based on electrochemistry have been applied to filtration systems to enhance the overall performance regarding water flux and fouling resistance. Self-cleaning was was first demonstrated in electrically conductive titanium meshes used as spacers for microfiltration (MF) (Abid, Lalia et al. 2017). The electric polarization of a spacer has also been utilized to enhance the performance of water filtration membranes. An electrically polarized titanium mesh was used for biofouling control in an RO system (Baek, Yoon et al. 2014). The effect of electrical polarization on the movement of ions in water and on membrane systems has been investigated by a variety of studies. Kim et al. introduced the effect of electric field on the mitigation of salt ions and showed that it also reduces the internal concentration polarization (ICP) (Kim, Ko et al. 2010). Son et al. investigated the ion migration induced by an electrical field on the water flux in an FO process by using a thin-film composite (TFC) membrane. This research showed that the electric field that brings protons closer to the membrane surface can provide enhanced osmotic pressure due to the high localized concentration of protons (Son, Kim et al. 2019). When functionalized graphene nano-sheets were used as a membrane, a faster movement of the salt ions under the electric field was also reported (Azamat 2016).

In this research, we propose an electrically-polarized graphene-blended (E-GRP) spacer as a salt adsorber prepared by 3D printing followed by a polarizing step (Fig. 1a and b). 3D printing is a versatile fabrication method, which allows to control the size, scale, and material of the final product (Koo, Ho et al. 2020, Yanar, Kallem et al. 2020, Yanar, Son et al. 2020, Kerdi, Qamar et al. 2021). As the ion movement (or concentration) and electric polarization on the draw side are crucially important in the FO process, the developed E-GRP spacer was employed on the draw side to investigate the changes in FO performance (Fig. 1c). We also tested it on the feed side with organic salts to check the performance to mitigate membrane scaling. To our knowledge, this is the first attempt to fabricate an electrically polarized conducting spacer and apply it to a membrane process.

**Materials and Methods**

The spacers were designed with the same filament thickness of 50 mil (1.27 mm) spacers with a pore size four times larger than that of commercial spacers (Fig. S1). The reason for the enlarged size parameters was to allow the membrane to have a larger osmotic concentration area and to avoid the precision limitations of Fused Deposition Modelling (FDM) 3D printers. A Computer Aided Design (CAD) model of the spacer with dimensions 45 mm × 60 mm × 1.905 mm was prepared using Autodesk Meshmixer and Blender (Fig. 2).

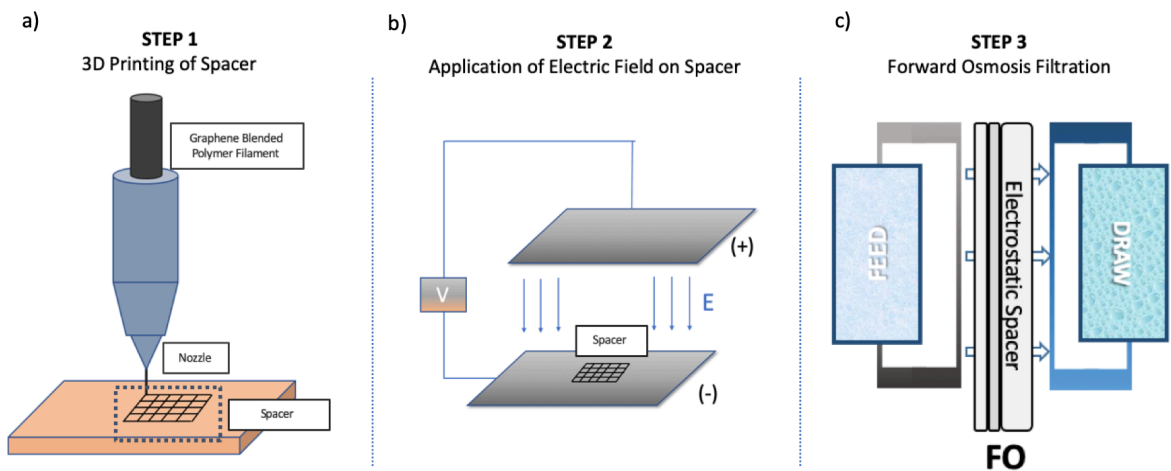

**Fig. 1** The fabrication and application processes of the electrostatic graphene-blended PLA spacers (E-GRP): a) 3D printing, b) electric polarization, and c) forward osmosis filtration

For the 3D printing of the samples, we utilized an FDM 3D printer (OpenCreators-Almond, Republic of Korea) (Fig. 1a), since it allows the freedom to choose the printing materials as opposed to other types of 3D printers. The fabricated PLA, graphene-blended PLA (GRP), and E-GRP were expected to have the same geometry, as the same printing procedure was applied. For the pristine PLA spacer, 1.75 mm of natural PLA filaments (PLABS, Republic of Korea) were used. For the GRP and E-GRP spacers, we utilized 1.75 mm of conductive graphene blended PLA filaments with an 8% graphene/PLA ratio and a volume resistivity of 0.6 Ohm·cm (Foster, Down et al. 2017, Carvalho Fernandes, Lynch et al. 2020) (Graphene Laboratories, United States). After the printing procedure, the spacers were washed first with ethyl alcohol and then with de-ionized (DI) water with the resistivity of 18.2 mΩ·cm at 25°C (Yanar, Son et al. 2020). An ultrasonicator (B8510-MT, Branson, USA) was used to remove the remaining filament particles from the spacers.

After the spacers were dried, one of the graphene-PLA spacers was placed into a lab-scale electric field system consisting of two parallel-plates at a distance of 20 cm. We subsequently applied ± 15 kV between the plates for 2 h to electrically polarize the GRP spacer to obtain an E-GRP one. The electric field was calculated with the following formula (Hu, Liang et al. 2020):

$$E = \frac{\Delta V}{d} \qquad (1)$$

where E is the electric field (V/cm), $\Delta V$ is the potential difference between the plates (V), and d is the distance between the plates (cm).

The surface charge densities of graphene spacers were also measured. Two flat graphene/PLA samples with an area of 3×1.5 cm² and a thickness of 200 μm were fabricated. One of these samples was electrically polarized for two hours in the same electrical field conditions above. Then, the zeta potentials of the samples were measured to calculate the surface charge densities using the Graham equation:

$$\sigma = \sqrt{8\varepsilon_r \varepsilon_0 kTI} \sinh(\frac{ze\zeta}{2kT}) \qquad (2)$$

where σ is the surface charge density, which is the hyperbolic function of the measured zeta potential $\zeta$. I is the total electrolyte concentration, $\varepsilon_0$ is the vacuum permittivity, which is 8.854×10⁻¹² F·m⁻¹, $\varepsilon_r$ is the relative dielectric permittivity of the solvent, $e$ is the elementary charge, equal to 1.602×10⁻¹⁹ C, kT is the thermal energy (k is the Boltzmann constant, which is 1.381×10⁻²³ J·K⁻¹ and T is the temperature in Kelvin), and z is the ion valence (Yang, Wu et al. 2017). For a 0.6 M NaCl solution, $\varepsilon_r$ can be calculated as follows:

$$\varepsilon_r = \varepsilon_w + 2\overline{\delta}c \tag{3}$$

where $\varepsilon_w$ is the dielectric permittivity of water, which is 78.2 at 25°C (Britannica) and $\overline{\delta}$ represents the relative contributions of the two ions in a single case ($\overline{\delta} = \frac{\delta^+ + \delta^-}{2}$). The value of $\delta^+_{Na^+}$ was taken as $-8$, $\delta^-_{Cl^-}$ as $-3$, which led to $\overline{\delta}$ being equal to $-5.5$. Finally, c is the concentration of the solution (Hasted, Ritson et al. 1948).

The Debye length in the NaCl electrolyte solution is also considered to observe the effect of electrolyte concentration for solid charge carrier's net electrostatic effect.

$$\lambda_D = \sqrt{\frac{\varepsilon_r \varepsilon_0 kT}{\Sigma_i \rho_{\infty i} e^2 z_i^2}} \tag{4}$$

where $\rho_{\infty i}$ is the number density of ion type i and $z_i$ is the ion valency (Smith, Lee et al. 2016).

The thickness of each spacer was first characterized at the millimeter scale by a scanning field emission electron microscope (FE-SEM, S-4700 Hitachi, Japan). The junction points of the spacers were also visualized to ensure that inter-membrane spaces were provided.

As we proposed this spacers for its ion attraction and for employing in a forward osmosis system, we applied it first on the draw side, where a highly concentrated sodium chloride (NaCl) solution is used. For comparison, the spacer was also tested on the feed side.

A forward osmosis system with an effective membrane area of 19.35 cm² and a total effective height of 2.6 mm was used in the cross-flow mode with a flow velocity of 200 cm³·min⁻¹ on both the feed and draw sides. A commercially available woven permeate spacer and commercial FO membranes extracted from an FO module (Toray Korea) were used to test the performance of the system (Fig. 1).

For the case where the driving force was the osmotic pressure across the membrane, a 0.6 M NaCl solution (draw solution) was paired with DI water (feed solution). The water weight and electrical conductivity on the feed side were automatically recorded by a computer every minute. Filtration tests were conducted for 1 h for each sample once the system was stabilized (i.e., stable water flux). The water flux, $J_w$, and the reverse solute flux, $J_s$, were calculated as given below:

$$J_w = \frac{V}{A_m \Delta t} \tag{5}$$

$$J_s = \frac{V_t C_t - V_0 C_0}{A_m t} \tag{6}$$

where $J_w$ is the water flux, $V$ is the volume of filtrated water (L), $A_m$ is the effective membrane area of the testing module, and $\Delta t$ is the permeation time (h) (Alayande, Obaid et al. 2019, Park, Yang et al. 2020). The reverse solute flux was obtained from the change in feed conductivity per minute and converted into gMH. $C_t$ (g·L$^{-1}$) and $V_t$ (L) are the concentration and volume of the feed solution measured at time t, respectively, and $C_0$ (g L$^{-1}$) and $V_0$ (L) are the initial concentration and volume of the feed solution, respectively. The concentration values were determined from the solution conductivities (Nguyen, Nguyen et al. 2015).

The performance of E-GRP as the feed spacer was further tested by placing a feed solution containing inorganic salts—19 mM NaCl, 20 mM sodium sulfate (Na$_2$SO$_4$), and 35 mM calcium chloride (CaCl$_2$)—and placing 0.6 M NaCl solution on the draw side. The solubility product of the Ca$^{2+}$ and SO$_4^{2-}$ concentrations in the feed solution was slightly higher (with a saturation index [SI] of 1.3), which can create gypsum scaling on the membrane at a reasonable flow rate, which was 200 cm$^3$·min$^{-1}$ for this experiment, as stated above (Mi and Elimelech 2010, Wang, Wang et al. 2016).

To further understand the effect of electric polarization on the ion attraction on the spacers, the spacers were each placed in 300 ml of 3 M NaCl solution for 12 h. This procedure was done right after the E-GRP polarization for 100 h. After a 12 h deposition of Na$^+$ and Cl$^-$ ions on the spacers, each was first placed in a 300 ml beaker filled with DI water and then in an ultrasonicator for 1 h. The conductivity of each solution with the released ions was subsequently measured after removing the spacers. Next, the conductivities were converted into total dissolved solids (ppm). The surface charge density in the 3 M NaCl solution was also calculated by taking $\varepsilon_r$ as 54 (Hasted, Ritson et al. 1948, Persson 2017).

**Results and Discussion**

The 3D printed samples were successfully fabricated with marginal differences from the CAD model (Fig. 2). The original design was 50 mil (1.27 mm) thick; however, the printed samples showed a relatively lower thickness due to the limited resolution of the layer by layer printing method of the FDM 3D printer. While the thicknesses of the GRP and E-GRP spacers were equal to 1.20 mm, that of the PLA spacer was 1.15 mm (< 4% difference). It can be noticed that blending graphene increased the rigidity of the melted filament, which resulted in a relatively higher precision than for the PLA spacer without blending (Fig. 3)

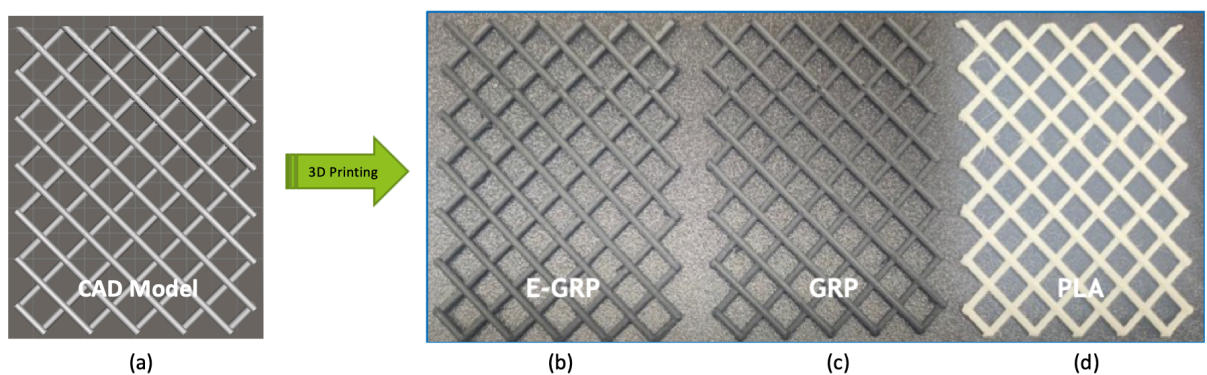

**Fig. 2:** a) The CAD model and the fabricated b) E-GRP, c) GRP, and d) PLA spacers.

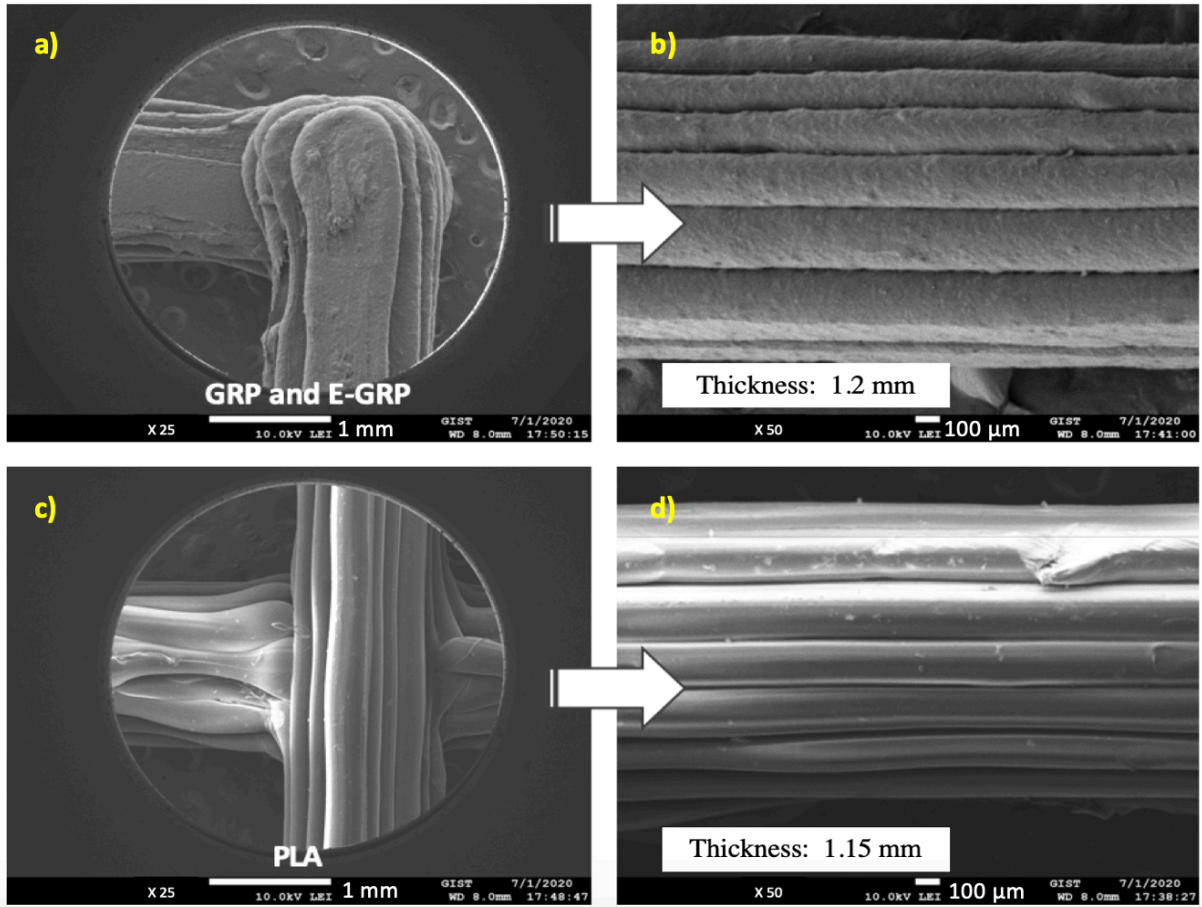

**Fig. 3:** Top and cross-sectional SEM images of the a, b) GRP and E-GRP and c, d) PLA spacers

After the E-GRP was polarized under a calculated electric field of 1.5 kV/cm, the prepared spacers were first tested by placing them on the draw side, as our purpose was to analyze the effect of ion attraction on the spacers. When the E-GRP spacer was used as a draw spacer, 32.4 ± 2 LMH of water flux (WF) and 13.4 ± 1.5 gMH of reverse solute flux (RSF) were measured. This measured WF was higher than that of the PLA (20.8 ± 2.1 LMH) or GRP (20.5 ± 2.3 LMH) spacers (Fig. 4a). Since the reverse solute flux is generally proportional to the water flux in the FO process, less solute permeation was observed for PLA (8.7 ± 1.2 gMH) or GRP (10.5 ± 2.1 gMH) compared to E-GRP. We explain this effect with the Gouy Chapman theory, which states that there should be exactly balanced by an equal and oppositely charged ions or counterions in the solution adjacent to the charged surface and a deficit of similarly charged ions or co-ions (Elimelech, Gregory et al. 1995).

Based on this, the interfacial potential at the polarized surface should increase the number of ions attached to it to an equal number of ions of opposite charge in the solution. Thus, the surface near to the spacer is expected to have a higher counter-ion concentration (Fig. S2), which will also increase the local osmotic concentration, resulting in higher flux. To further understand this phenomenon, we calculated the surface charge density on the E-GRP and GRP spacers as a hyperbolic function of the surface zeta potential using the Grahame equation. As we were not able to measure the surface zeta potential of the spacers due their shapes, we prepared 200 μm flat samples from the same printing material and polarized one of them in the same conditions as the E-GRP spacer. From the measured zeta potentials, the calculated surface charge density of the E-GRP (−4.2 mC/m$^2$) spacer was found to be higher than that of the GRP (−0.27 mC/m$^2$) one (Fig. S3). Therefore, the surface charge density was enhanced through the polarization of the graphene-blended spacer, an effect that could provide higher local osmotic concentration at the membrane surfaces near the polarized spacers. Furthermore, the collection of ions on the spacer reduced the external concentration polarization on the surface of the membrane.

The effect of using E-GRP as a feed spacer was also investigated (Fig. 4b). On the feed side, E-GRP (WF: 28.8 ± 0.6 LMH and RSF: 14.7 ± 4.4 gMH) had a performance similar to GRP (WF: 25.1 ± 0.8 LMH and RSF: 15.8 ± 2.25 gMH) and PLA (WF: 27.1 ± 0.5 LMH and RSF: 15 ± 3.8 gMH). As there were no ions present in the feed solution, the E-GRP spacer was not able to adsorb ions; thereby, only marginal changes in the water flux and reverse solute flux were found (Pal 2017).

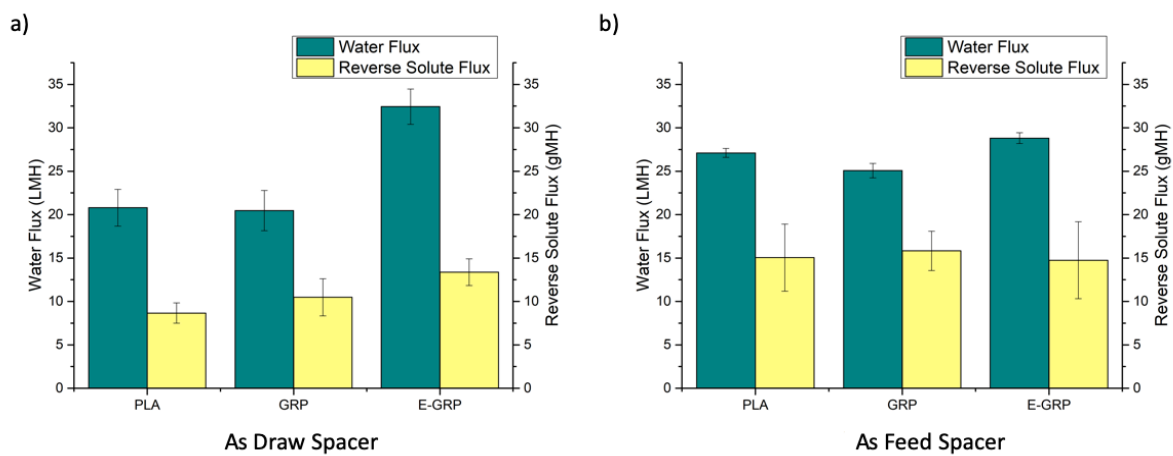

**Fig. 4**: The pure water flux of each spacer on the a) draw side and b) feed side

Ion adsorption and desorption capacities of each spacer were further investigated by measuring the amount of collected ions on E-GRP, GRP and PLA by dipping in 3M NaCl solutions (Fig. S4). E-GRP clearly showed the higher amount of collected ions on it compared to PLA and GRP spacers. E-GRP collected 129,000 ppm ions on it during 12 h dipping right after the polarizing, whereas it decreased to 106,000 ppm 100 h after polarization. In contrast, GRP had in the range of 92,000-93,000 ppm, while PLA had in the range of 73,000-75,000 ppm for both tests which are done right after and 100 h after polarizing (Fig. 5). Thus, the effect of electrostatic forces on the collection of ions on spacers instead of the membranes could be expected. In order to support this data, surface charge densities of E-GRP (-7.4 C/m$^2$) and GRP (-0.53 C/m$^2$) were also calculated for 3 M NaCl solution from Grahame Equation (Fig. S5). It shows that surface charge densities are increasing with the increasing solvent concentration, which also effect total collected ions on spacer surfaces. Therefore, E-GRP's superior ion adsorption capacity was proportionally increased to the solvent concentration. Although relatively thin Debye length of 4.61 nm for 3 M NaCl was calculated, it was also previously studied that the Debye length is not applicable for higher solvent concentrations of >0.1 M as it does not also consider the effect of surface charges of the solid surface (Conway, Bockris et al. 1951, Philpott and Glosli 1995).

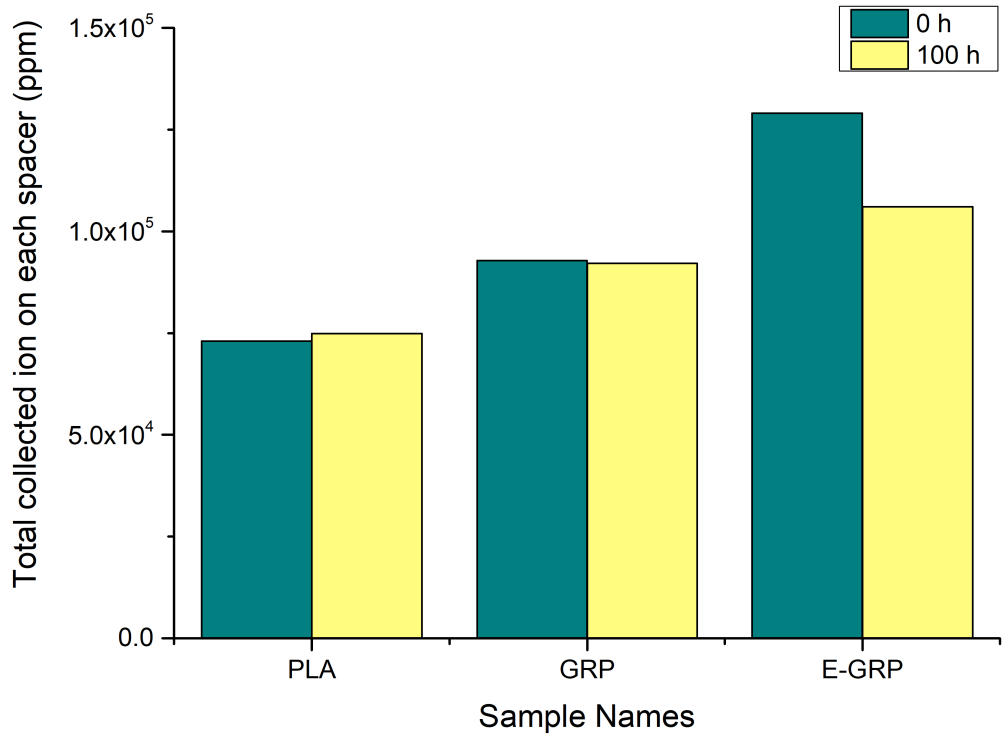

**Fig. 5** Total collected ion on each spacer right after (green bars) and 100 h (yellow bars) after electric field application to E-GRP

E-GRP as a feed spacer was further tested to observe the scaling performance, when the feed was concentrated with inorganic salts of NaCl, $Na_2SO_4$, and $CaCl_2$. The use of the E-GRP as a feed spacer has suppressed the gypsum scaling on the membrane surface at the feed side. Only ~20% of flux reduction was found for E-GRP, whereas >80% of water flux was declined for the pristine GRP during 4 hours of operation (Fig. 6). This improved gypsum scaling resistance of the E-GRP spacer could be attributed to the electrostatic forces between the gypsum aggregation and membrane surface. For instance, collection of cations on polarized spacer surface could disrupt the gypsum formation on membrane surface because the electrostatic forces are well known to effectively disperse gypsum aggregation (Cheng, Wang et al. 2016). Furthermore, some gypsum could also be attracted by the polarized spacer instead of blocking the membrane surface, as it is known from the literature that there is an enhanced ionic interaction between the negatively charged surfaces (E-GRP) and gypsum particles (Chen, Su et al. 2013). This was further supported by SEM images as less gypsum formation was observed for E-GRP compared to that of GRP (Fig. 7).

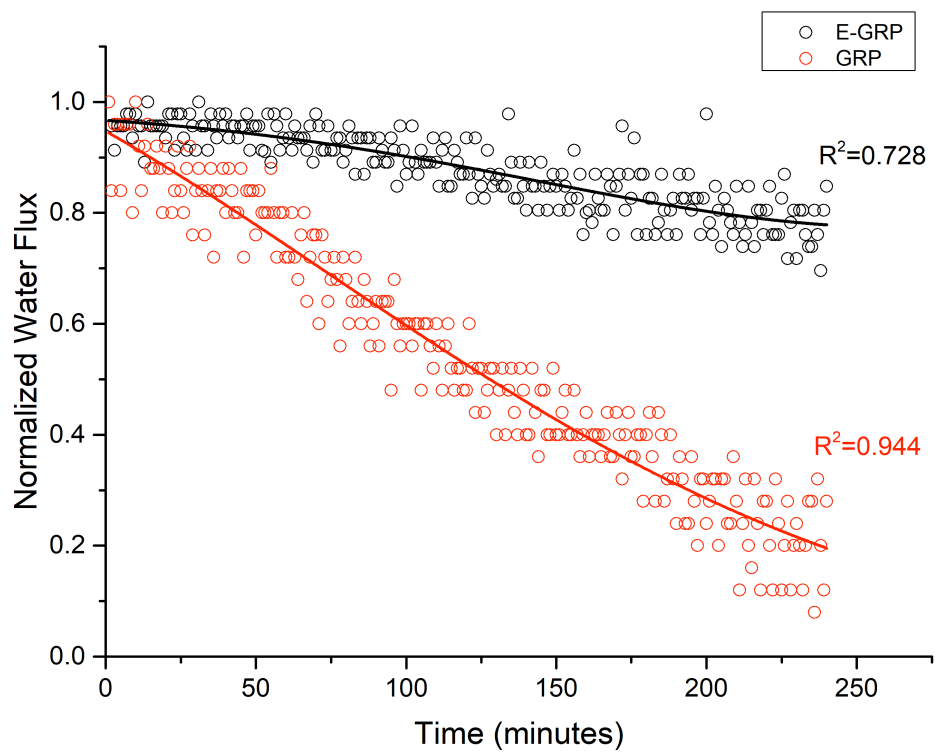

**Fig. 6** Normalized Water Flux for 4 hours gypsum scaling of membranes with GRP and E-GRP

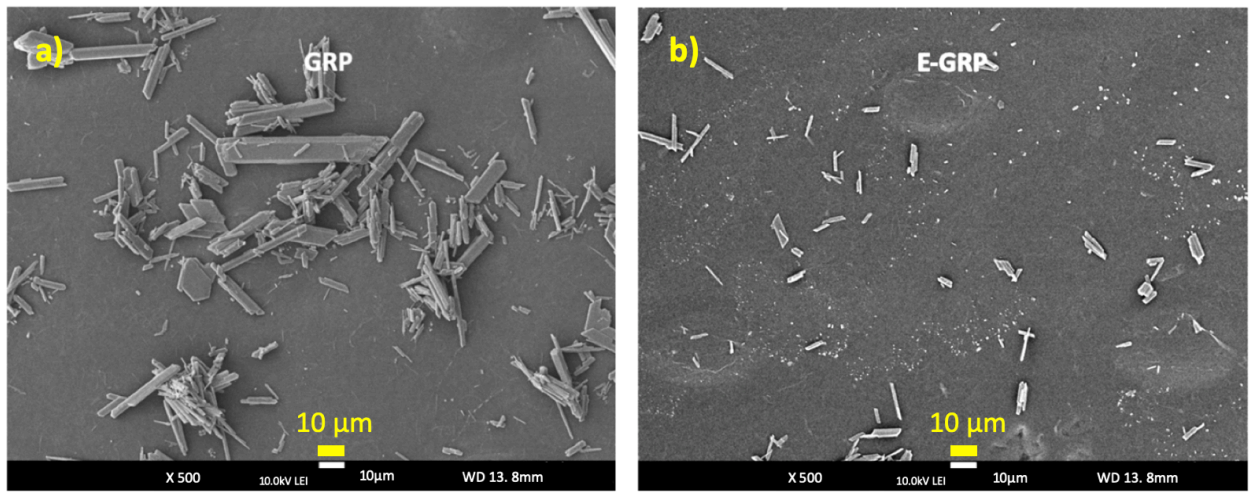

**Fig 7.** Gypsum scaled membranes of a) E-GRP and b) GRP

**Conclusion**

In this research, we propose a novel E-GRP spacer to maximize water flux and minimize gypsum scaling in FO application. The developed E-GRP has collected ions on itself and thereby enhanced local osmotic concentration near the membrane surface. When electrically polarized graphene blended polymer spacer E-GRP was used as draw spacer, it showed over 50% higher water flux compared to the pristine non-conducting spacer (PLA) and graphene-blended spacer (GRP). When E-GRP was used as a feed spacer, it did not show highly enhanced performance due to the absence of ion present in the feed solution. However, when scaling test was performed by dissolving inorganic salts in feed side, E-GRP showed a great performance to mitigate membrane scaling by disrupting gypsum formation on membrane surface as result of collection of gypsum forming ions on spacer surface rather than membrane active layer and its effect on the dispersion of gypsum aggregation.

To make this approach more feasible in practical application, periodic polarization (or self-polarization) of spacers should be further investigated. One suggestion could be the use of E-GRP in self-polarizing systems by connecting external circuit. In addition, other types of fouling such as organic and biofouling should be also tested by considering larger scale applications.

**Acknowledgements**

This work was supported by National Research Foundation of Korea (NRF) projects funded by Korean Government (MSIT) with the grant numbers of No. 2020R1A2C2010808 and No. 2020M3H5A1081105.

a) Cross Section View

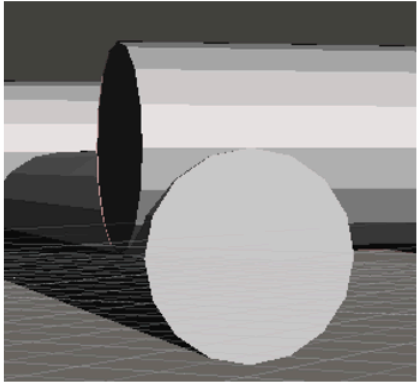

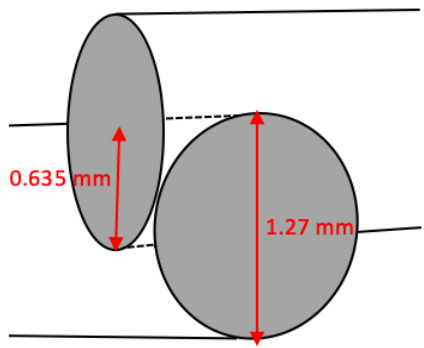

b) Front View

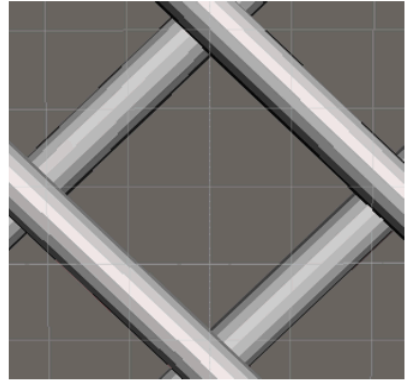

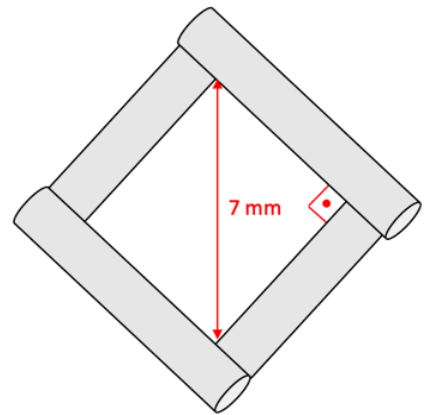

**Fig. S1.** a) Cross-section and b) front views of the spacer designs (CAD models and illustrative drawings)

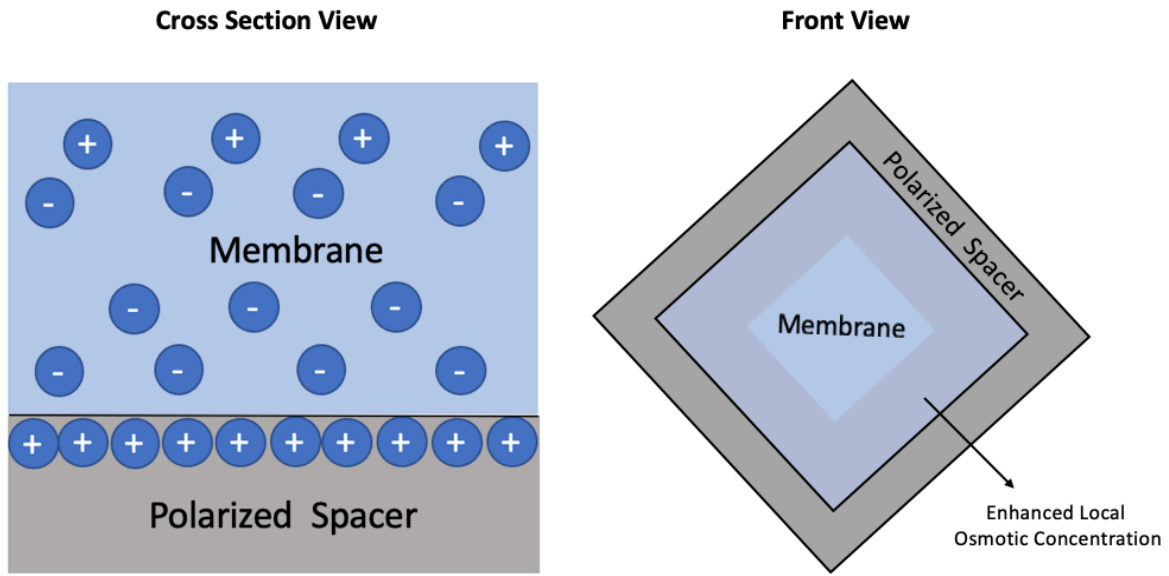

**Fig. S2.** a) Cross-section and b) front views of the enhancement of the local osmotic concentration near surface of membrane to spacer

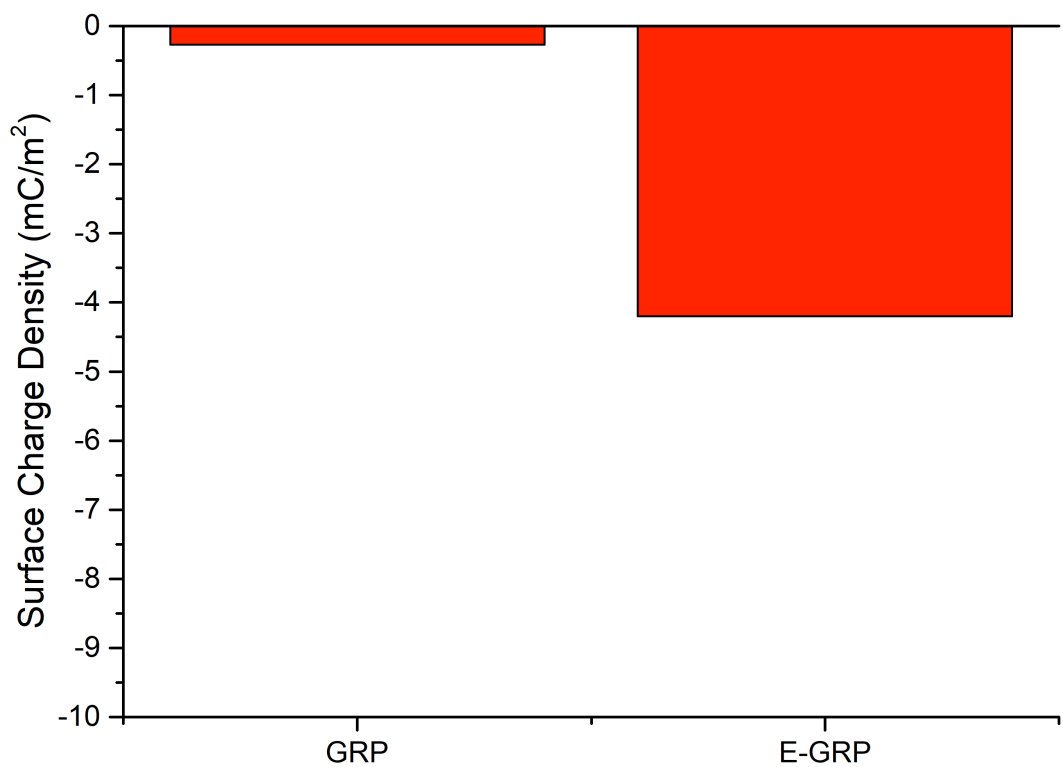

**Fig. S3.** Surface charge densities of the GRP and E-GRP samples in the 0.6 M NaCl solution

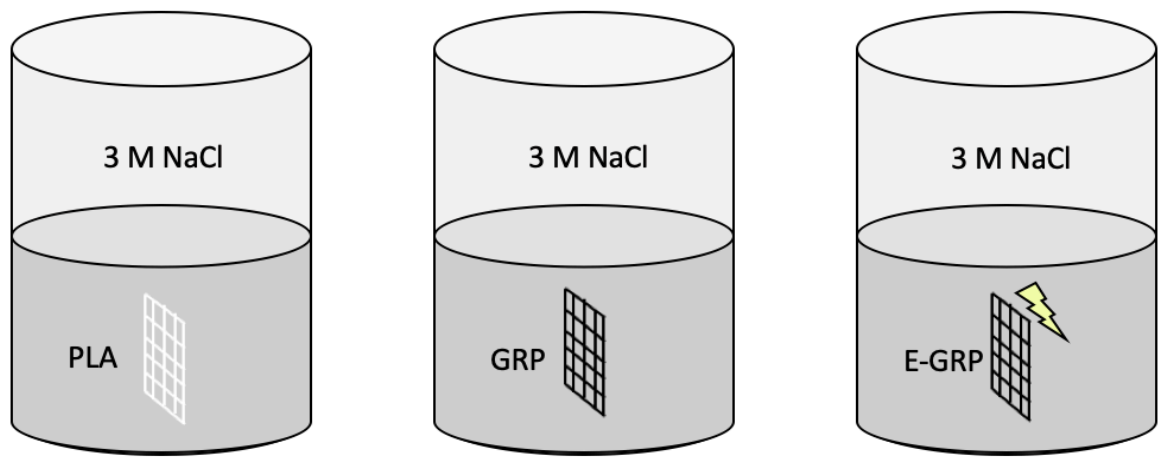

**Fig. S4.** The PLA, GRP, and E-GRP spacers were submerged for 12 h in a 3 M NaCl solution to observe their ion absorption performance

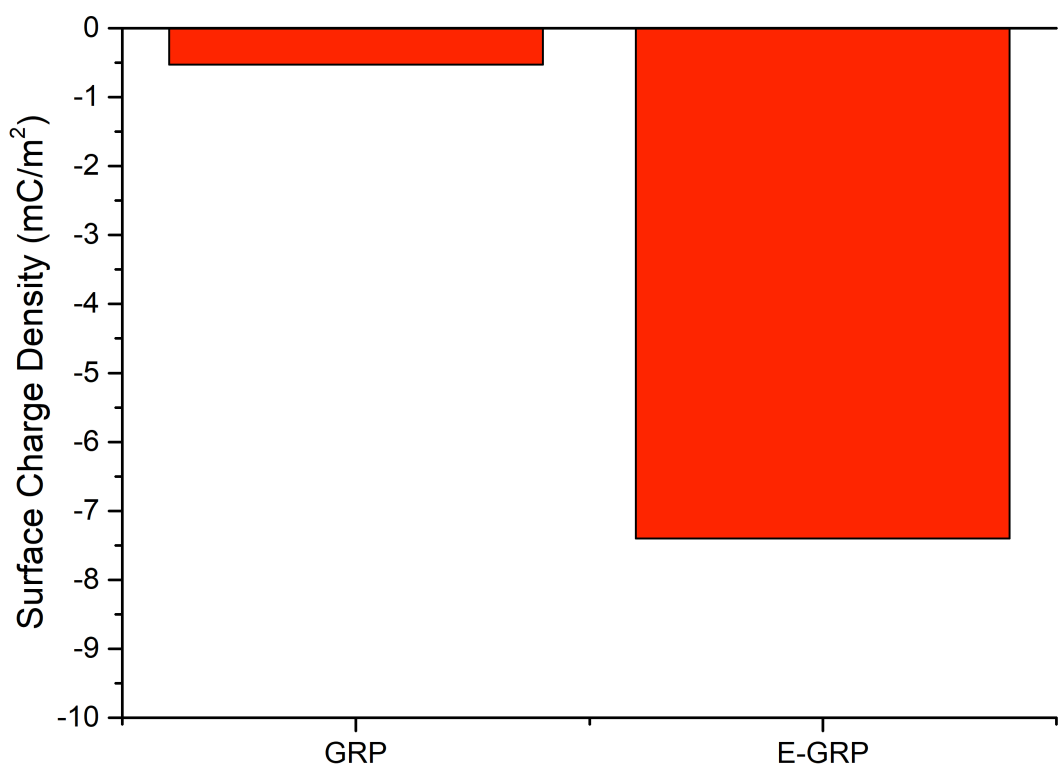

**Fig. S5.** Surface charge densities of the GRP and E-GRP spacers in a 3 M NaCl solution